\documentstyle[aps,twocolumn,prb,epsf]{revtex}
\begin{document}
\draft
\twocolumn[\hsize\textwidth\columnwidth\hsize\csname @twocolumnfalse\endcsname
\title{Electromigration of
Single-Layer Clusters}
\author{O.\ Pierre-Louis$^{*,\dagger}$ and T.\  L.\ Einstein$^{\dagger}$}
\address{$^*$LSP Grephe, CNRS, UJF-Grenoble 1,
BP87, F38402 Saint Martin d'H\`eres, France \\
$^\dagger$Department of Physics,
University of Maryland, College Park, MD
20742-4111}

\maketitle
\begin{abstract}
Single-layer atom or vacancy clusters in the presence of electromigration
are studied theoretically assuming an isotropic medium.
A variety of distinctive behaviors distinguish the response in the three
standard
limiting cases of periphery diffusion (PD), terrace diffusion (TD),
and evaporation-condensation (EC). A general model provides power laws
describing the
size dependence of the drift velocity in these limits, consistent
with established results in the case of PD. The validity of the widely used
quasistatic limit is calculated. Atom and vacancy clusters drift in opposite
directions in the PD limit but in the same direction otherwise. 
In absence of PD, linear
stability analysis reveals a new type of morphological instability,
not leading to island break-down. For strong electromigration,
Monte Carlo simulations show that
clusters then destabilize into slits, in
contrast to splitting in the PD limit. Electromigration affects the diffusion
coefficient of
the cluster and morphological fluctuations, the latter diverging
at the instability
threshold. An instrinsic attachment-detachment bias displays the same scaling
signature as PD in the drift velocity.
\end{abstract}
\pacs{olivier.pierre-louis@ujf-grenoble.fr; einstein@physics.umd.edu}
]

\section{Introduction}
Recognized as a key source of size limitation in electrical
devices, surface electromigration has important consequences for
surface morphology. Latyshev {\it et al}.\ \cite{lask} discovered
that it can induce step bunching on Si(111) vicinal surfaces. How
electromigration affects more complex surface structures is still
poorly understood.  In this paper we explore the effect of electromigration
of atoms or vacancies on single-layer clusters.
Responses
of these clusters to electromigration share similarities with void
behavior in metallic electric lines\cite{Ho} and
with electromigration of oxygen disordered domains in YBaCuO thin
films\cite{mlb,ws} and lies
in the active area of driven diffusive systems.\cite{sz}

In addition to the intrinsic interest in how island properties
are affected by the symmetry-breaking electromigration force,
we will point out further mesoscopic ways to address current
controversies about the dominant mechanism of mass transport.\cite{Rosen}
Moreover, these properties are needed to
model surfaces undergoing electromigration at larger scales.  For
example, electromigration-induced coalescence models should use
non-equilibrium diffusion constants and steady-state drift
velocities.

In the presence of an electric current,
the electromigration force is usually described as
${\bf F} = z^*e{\bf E}$, where $e$ is the magnitude of the electron
charge and
$z^*$ is an effective valence, generally non-integer, which takes
into account both the electrostatic interaction between
the electric field {\bf E} and the
charge distribution on the affected atoms (``direct" force) and the
frictional force resulting from the transfer of momentum from the
charge carriers to these atoms (``wind" force).  The
long-standing controversies regarding bulk electromigration
highlight the complexity of microscopic calculations, and
particularly how conduction electrons may screen the direct charge
of the atoms.\cite{controversy3D}
Non-trivial aspects of the flow of the electron cloud
about the diffusion path might also contribute to the direct
effect on the surface.\cite{kk}  For metals, the
electromigration effect is primarily due to wind force;
calculated values include $z^*
\approx -30$ for Al on semi-infinite jellium \cite{Ish} and $z^*
\approx -21$ for Cu on Cu \{111\}.\cite{rew}  (Near a step edge, one can
expect quantitative but not order-of-magnitude changes in $z^*$; e.g. for an
atom {\it in} a close-packed step edge on Al\{001\}, $z^*
\approx -43.$\cite{Rous}) For semiconductors, both forces are small,
with a resultant $|z^*|
\approx 0.001 - 0.1$.\cite{sc/}
A characteristic length can be associated with electromigration:
$\xi \equiv k_BT/F$. In typical experimental conditions,
$\xi \sim 10^{8}$ atomic spacings for Si and $\xi \sim 10^{5}$ for metals.
We base our study of the equilibrium
behavior of these islands on the framework of the continuum theory of
Khare {\it et al}.\cite{ke} We consider electromigration to be a
perturbation inducing a macroscopic current,
as done previously in Refs. \onlinecite{Ho} and \onlinecite{stoy}.
This approximation is justified because $\xi$ is much
larger than the atomic spacing. The adatom flux on terraces
is then equal to $c \langle v \rangle$.
The mean drift
velocity $\langle v \rangle$ of an adatom due to electromigration
is calculated from the Einstein relation:
\begin{eqnarray}
\langle v \rangle = D {F \over k_BT} = {D \over \xi} \; .
\label{e:einstein_1}
\end{eqnarray}

Single-vacancy motion
may dominate over atom motion in the mass transport on some surfaces.  (There
is evidence of this on Cu \{001\}.\cite{hkgibh,hkma})  In that case one must
use an effective charge $z^*$ appropriate for vacancies and reverse the
signs of
step curvatures in computing restoring forces due to line tension.  The
resulting
modifications are relatively straightforward but tend to muddle the subsequent
descriptions.  Accordingly, we do not explicitly consider vacancy transport
below.

In the notation of Ref. \onlinecite{ke}, there are three
limiting mass transport modes:
periphery diffusion (PD), terrace diffusion (TD), and
(two-dimensional) evaporation-condensation (EC) or
attachment-detachment (see Fig.~1).
In particular, there is a distinctive size
dependence of the tracer diffusion constant $D_c^{eq}$ of a cluster
at thermodynamic equilibrium,
with characteristic exponent $\alpha$ =3, 2, and 1, respectively:
\begin{equation}
D_c^{eq}
 \sim R_0^{-\alpha}\ ,
\end{equation}
\noindent where $R_0$ is its average radius.
Intermediate values can also be obtained in restricted regions of
parameter space as one crosses over from one limiting regime to another.
As a first qualitative approach, the mean cluster
drift velocity should be given by the Einstein
relation  in a way very similar to the drift velocity
of one adatom (Eq.\ (\ref{e:einstein_1})):
it is equal to the cluster mobility
(or equilibrium diffusion constant $D_c^{eq}$)
time the force exerted on the whole cluster
$\overline{F} = \pi R_0^2F/a^2$, and divided
by $k_BT$:
\begin{equation}
|\overline{V}|\approx  D_c^{eq}{\overline{F} \over k_BT}
=  D_c^{eq}{\pi R_0^2 \over a^2 \xi} \sim R_0^{2-\alpha} \; .
\label{e:Vscale}
\end{equation}
This scaling law will be confirmed later.
The intermediate expressions are in fact exact
for models in which atoms approach and
leave the island only from the lower edge
(one-sided step), or in the PD regime.
(In the more general case, some subtleties arise.)

We further investigate the shape changes associated with
steady states. Circular steady states are found in the PD and TD regimes.
But when attachment and detachment of atoms to the steps is not instantaneous,
the cluster elongates. This can be understood
by analogy to the Bernoulli effect. Stability of these steady states
is studied. Besides the expected splitting of clusters for PD,
we find another morphological instability in the TD case. Monte Carlo
simulations indicate that it leads to
slit formation.
Fluctuations and the non-equilibrium diffusion constant are also calculated
within the framework of a Langevin model,
and appear to be corrected by a bias term
proportional to $1/\xi^2$ for weak electromigration.

\section{Model}

Our formal approach is inspired by a previous examination of
non-equilibrium step
meandering on vicinal surfaces.\cite{oplcm} For simplicity and to avoid
distracting
complications, we take all step and terrace properties to be isotropic. We
forbid
adsorption onto or desorption from the terrace. Evolution of the mobile-atom
concentration on terraces is then given by the following conservation law:

\begin{eqnarray}
\partial_tc & = &
-\nabla \cdot {\bf J},
\label{e:mc_ter}
\\
{\bf J} & \equiv & -D\nabla c + {D \over \xi}c\ \hat{\bf x} +{\bf q} \; .
\label{e:dtc}
\end{eqnarray}

\noindent
${\bf J}$ is the mass flux on terraces, and $D$ is likewise the
adatom diffusion constant on terraces. The correlations  of the conserved
noise {\bf q} are given below. In the widely used quasistatic limit, one
considers that the adatom concentration reaches a steady state on time scales
much smaller than that of step motion. Then
\begin{eqnarray}
\nabla \cdot {\bf J}=0
\label{e:mc_quas}
\end{eqnarray}
has to be solved on terraces instead of Eq.(\ref{e:mc_ter}).
Adatom electromigration is taken to be unaffected
by the presence of steps; thus, $\xi$ is uniform. Atom
exchange between the step and the 2D ``gas" on the terrace is proportional
to the
deviation from equilibrium, expressed in the linear kinetic relation:

\begin{equation}
{\bf \hat{n}}\cdot {\bf J}_\pm = \mp \nu_{\pm}(c - c_{eq} -\eta_\pm).
\label{e:lin_kin}
\end{equation}

\noindent In our notation, + or $-$ denote the {\it lower} or the {\it
upper} side,
respectively, of the step forming the boundary of the island.  Thus, for
atom islands, + is the exterior while for vacancy islands it is the
interior.  The unit vector ${\bf \hat{n}}$ normal to the step
points toward the $+$ side. The kinetic coefficients
$\nu_{\pm}$ describe attachment and
detachment at the two sides of the step edge, and $\eta_{\pm}$ are
non-conserved noises.  Attachment lengths are defined by
$d_{\pm} \equiv D/\nu_{\pm}$; the larger $d_{\pm}$, the smaller the
chance that
an atom will detach from or attach to the step. When
$d_{\pm}$
are small, the dynamics are diffusion limited (TD).  For large but finite
$d_{\pm}$, atoms attach only after a large number of trials, and dynamics are
thus limited by attachment and detachment (EC).  In the limit that
$d_-\rightarrow
\infty$, the model reduces to an exterior (or interior) model for atom (or
vacancy) islands, because atom exchange occurs on only one (viz., the lower)
side of the step. The equilibrium concentration $c_{eq}$ must include
corrections due to boundary curvature, as given by the Gibbs-Thomson relation
\begin{eqnarray}
c_{eq} =
c_{eq}^0\exp(\Gamma\kappa) \;,
\end{eqnarray}
where $c_{eq}^0$ is the equilibrium concentration in the vinicity
of a traight step,
$\kappa$ is the step curvature (counted positive for a convex step
cf. Eq.\ (\ref{e:curv_pol})),
and $\Gamma$ is the capillary length:
\begin{eqnarray}
\Gamma=a^2\tilde{\beta}/k_BT \, .
\end{eqnarray}
The step stiffness $\tilde{\beta}$ is taken to be isotropic, as noted above;
$a$ is the lattice constant, and $T$ is the temperature.
Finally, the normal velocity $V_n$ (i.e., along ${\bf \hat{n}}$) of the
step is given by mass conservation

\begin{eqnarray}
V_n  =  a^2{\bf \hat{n}}\cdot \big( {\bf J}\big|_+
 - {\bf J}\big|_- \big)
- \partial_sJ_{st} \, ,
\label{e:conslaw}
\end{eqnarray}
where $s$ is the arclength along the step.
We consider the low-concentration limit,
where $c \ll 1/a^2$, i.e.
the atom concentration is taken to be much lower than in the bulk.
Hence, the second term, representing advection, in
${\bf J}|_\pm=-D\nabla c_\pm+{\bf \hat{n}}V_n c_\pm$ is neglected.
$J_{st}$ is the mass flux along the step.
In addition to the usual relaxation term related to chemical potential
gradient, electromigration induces a new term
$D_LF_{st}n_y/k_BT=D_Ln_y/\xi_{st}$, with
$\xi_{st}\equiv k_BT/F_{st}$.
$F_{st}$ is the force exerted on mobile atoms at
the step edge, and $n_y$ is the $y$ component of ${\bf \hat{n}}$. Thus, we have
\begin{eqnarray}
J_{st}=
a{D_L\over \xi_{st}}\partial_s[R(\theta)\cos(\theta)]
-a D_L\partial_s(\Gamma\kappa) - q_{st},
\label{e:Vn}
\end{eqnarray}
where $q_{st}$ is a Langevin force.
$R(\theta)$ is the distance from the center of
mass of the cluster, and
$\theta$ is the polar angle with respect to the $x$ axis
(see Fig.1).
The macroscopic step diffusion constant could be defined as
$D_L \equiv a^2c_{st}D_{st}$, where $c_{st}$ is
concentration of mobile edge-atoms, and $D_{st}$ is the diffusion constant
for motion of these atoms along the step, or using
the Kubo formula. \cite{vil_pimp}
In the following,
we will take $D_L$ to be uniform along the step and
constant in time.

In a local thermodynamical equilibrium approximation, noise correlations
are written
\begin{eqnarray}
\langle q_i(x,y,t)q_j(x^{\prime},y^{\prime},t^{\prime})\rangle =
2Dc(x,y,t)\delta \, ,
\nonumber \\
\langle \eta_\pm(s,t) \eta_\pm(s^{\prime},t^{\prime})\rangle
=\frac{2c(s,t)|_\pm}{\nu_\pm}\delta(s-s^{\prime})\delta(t-t^{\prime})\, ,
\nonumber\\
\langle q_{st}(s,t)q_{st}(s^{\prime},t^{\prime})\rangle =
\frac{a^3D_L}{\pi}
\delta(s-s^{\prime})\delta(t-t^{\prime})\, ,
\end{eqnarray}

\noindent where $\delta \equiv
\delta(x-x^{\prime})\delta(y-y^{\prime})\delta(t-t^{\prime})\delta_{ij}$,
and $c(s,t)|_\pm$ is the adatom concentration on the terraces in the
vicinity of the island edge.

In describing the fluctuations of these islands, we consider only
the case in which overhangs can be ignored. Thus, the periphery
can be described in polar coordinates by $R(\theta)$.
For small fluctuations from
the circular, $R(\theta)= R_0 +\rho(\theta)$.  The Fourier transform
of $\rho$ is given by
\begin{eqnarray}
\rho_{n\omega} = \int_0^{2\pi}
{d\theta \over 2\pi} \int_{-\infty}^{+\infty}
{dt \over 2 \pi}
\rho(\theta,t) e^{-in\theta-i\omega t}.
\label{e:fourier}
\end{eqnarray}
The equilibrium
properties of such clusters (for $F$=0) were calculated
in detail by Khare {\it et al}. \cite{ke}
They are able to evaluate the equilibrium diffusion constant
under general conditions.  Equipartition of energy provides
the static spectrum:\cite{kex}
\begin{equation}
\langle |\rho_n|^2 \rangle_{\rm eq} =
{ k_BTR_0 \over 2\pi(n^2-1)\tilde{\beta}} \; .
\label{e:stat_spec}
\end{equation}
The divergence of $\langle |\rho_1|^2 \rangle_{\rm eq}$ reflects
the absence of an energy cost to translate the whole island: this
is a ``Goldstone mode."
The mode $n=0$ is forbidden due to area
conservation for PD dynamics. But generally in the presence of non-conserved
dynamics, the system will tend to minimize step length, and clusters
are unstable: they expand or shrink, depending on the precise kinetics
and environment. As an example, an assembly of clusters will coarsen due to
Ostwald\cite{ost} or coalescence (``Smoluchowski")\cite{smol}
ripening.

Let ${\bf r}_{\rm CM}$ denote the position of the center of mass of the
cluster. We define the mean velocity:
\begin{eqnarray}
\overline{\bf V}=\langle\partial_t{\bf r}_{\rm CM}\rangle,
\end{eqnarray}
and the non-equilibrium tracer diffusion constant:
\begin{equation}
D_c={\langle |{\bf r}_{\rm CM}-\overline{\bf V}t|^2 \rangle \over 4t}\, ,
\end{equation}
which reduces to the usual equilibrium cluster diffusion constant
$D_c^{eq}$ when $\overline{\bf V}=0$.
This quantity could also be defined as half the
diffusion constant describing the relative motion of two identical
islands, whether they are drifting or not. This latter
definition has been useful experimentally.\cite{mrpc}
Shape fluctuations of the cluster are measured by the island
roughness
\begin{equation}
W^2 \equiv \langle R^2 \rangle - \langle R \rangle^2.
\end{equation}
At equilibrium, using Eq.\ (\ref{e:stat_spec}), we find:
\begin{equation}
W^2_{eq}={ 3k_BTR_0 \over 4\tilde{\beta}} \, .
\end{equation}
As a static property, $W^2_{eq}$ does not depend on the transport mechanism,
just like the static spectrum in Eq.(\ref{e:stat_spec}).

\begin{figure}[htb]
\epsfxsize=6cm
\epsfysize=8cm
\vspace{3 cm}
\centerline{\epsfbox[50 -100 500 500]{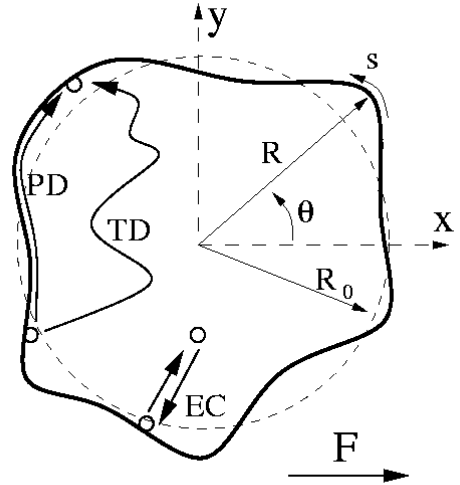}}
\vspace{-3 cm}
\caption{\it\protect \small
Three processes are involved in mass transport:
Diffusion along the step, diffusion across the terraces, and
attachment-detachment of adatoms at the step. When one of these processes
is slower than the others, we get the
PD, TD, or EC regime, respectively.
Electromigration force $F$ is taken
to be along the $x$ axis in the text.
$R$ is the distance from the center of the island,
$\theta$ is the polar angle with respect to the $x$ axis,
and $s$ is the arclength along the step.
$R_0$ is the mean radius.
}
\end{figure}

A major goal of our analysis is to explore properties
of driven clusters on the dominant
mechanism of mass transport.
A crude criterion for being in the PD regime
follows from checking wether the diffusion constant is dominated by
line diffusion or diffusion across the terrace.
Comparing expressions
(\ref{e:dc_pd}) and (\ref{e:dc_td}) for $D_c^{eq}$ given below, one finds:
$R_0^2 \ll (R_0+d)\ell_c$,
where $d=d_+$ or $d_-$ and $\ell_c\equiv D_L/Dac_{eq}^0$.\cite{note_lc}
The opposite case, $R_0^2 \gg (R_0+d)\ell_c $,
will be denoted as the 2D transport regime.
In this regime, TD and EC limits
are defined by $R_0 \gg d_+$ or $d_-$,
and $R_0 \ll d_+$ and $d_-$, respectively.
Note that the cluster radius $R_0$ is involved in these relations:
it is a geometrical cut-off for long-wavelength fluctuations.
A more general definition of these regimes involving wavelength
dependence can be found in Ref. \onlinecite{ke}.

\section{PD limit}
\subsection{Steady states and stability analysis}

Some known results about 2D voids in metal electric line can be transposed
directly to the case of vacancy islands in the PD limit.
Best known is ``Ho's law," which
states that the drift velocity varies inversely with the radius.\cite{Ho,grub}
Behavior consistent with this relationship was seen in recent Monte Carlo
simulations of electromigration on Cu\{001\} using semi-empirical energy
barriers.\cite{mbmk}
In our formalism, the drift velocity (in the $x$-direction) of a 2D cluster
with mean radius
$R_0$ is:

\begin{equation}
\overline{V} =\phi \frac{aD_L}{\xi_{st} R_0} \; ,
\label{e:Ho}
\end{equation}

\noindent $\phi$ = 1 ($-1$) denotes atom (vacancy)
islands.
Thus, atom islands and vacancy clusters drift in
opposite directions.  With current densities about $10^4$--$10^6$
A cm$^{-2}$ at 600 K on metals, we find $\xi_{st} \approx
10^5$\AA. Assuming next an activation barrier of $\sim$0.5eV along
the periphery and a concentration of mobile atoms per site $c_{st}
\approx 0.1$, we estimate $D_L \approx 10^9$\AA$^2$/s; thus, for an
island of radius 10$^3$\AA, $\overline{V} \approx$  10\AA/s.

\begin{figure}
\centerline{
\epsfxsize=2.5in
\epsfbox{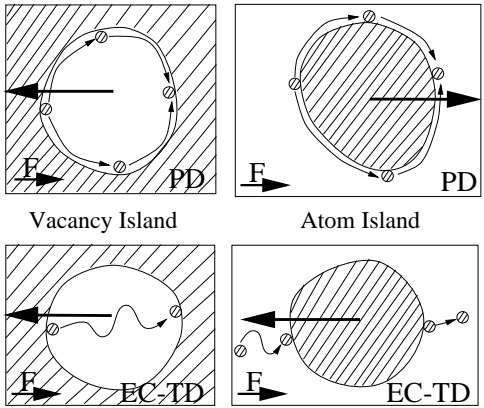}}
\vspace{0.5 cm}
\caption{\it\protect \small
Drift direction of island as a function
of mass transport mechanism. We have assumed a
direct Ehrlich-Schwoebel effect.
}
\end{figure}

Stability analysis for voids \cite{mb,wsh} can also be applied to our case.
Since we assume {\it isotropic step properties}, the equilibrium shape
of the island is circular.  With increasing electromigration, this
steady-state shape persists so long as the island is stable.
A hand-waving argument allows us to retrieve the instability
threshold found in Refs. \onlinecite{mb} and \onlinecite{wsh}.
We simply have to compare fluxes along the step close
to a protrusion of the step inside the cluster.
The flux contributing to destabilization is $J_{destab}\sim D_L/\xi$.
Stabilizing effects related to line tension effect result in
a flux $J_{stab}\sim D_L\Gamma/R_0^2$, where curvature changes
are taken to be $\sim 1/R_0$ between points separated
by a distance $\sim R_0$. The protrusion increases if
$J_{destab}>J_{stab}$.
The instability criterion is then $|\chi| > \chi_c$,
where $\chi \equiv R_0^2/\Gamma \xi =
(F_{\rm em}/\tilde{\beta})(R_0/a)^2$.
According to a linear stability analysis, $\chi_c=10.65$.\cite{mb,wsh}
With the preceding
parameters and $\tilde{\beta}$ =0.3eV/$a$, we find that islands with
radius larger than $R_{c} \approx 5\cdot10^3$\AA \ will be
unstable. The instability appears at a characteristic time $\tau =
\chi_c\Gamma \xi^2/(aD_L) \approx 10^2$s.
Thus, this phenomenon should be observable
experimentally.  The instability criterion
is the same for atom and vacancy islands.

Since electromigration of single-layer clusters is
unaffected by their shape, our voids can be characterized as
{\it conducting}. In contrast, voids in electric lines are
essentially {\it insulating}.  The resulting current-crowding effects
qualitatively
change the nature of the instability: the void is linearly stable, but becomes
unstable under finite perturbations.  At late stages of this instability,
the void splits.\cite{Schim}

\subsection{Diffusion constant and fluctuations}
Keeping only the line
diffusion term in Eq. (\ref{e:conslaw}),    
one gets in the frame moving with the mean velocity
$\overline{V}$ of the island:
\begin{eqnarray}
&& V_n-\overline{V}n_x = -\partial_s[J_{st}]
\nonumber \\
&=& -\partial_s \left[{a \over \xi_{st}}
D_L\partial_s[R(\theta) \cos (\theta)]
-a D_L\partial_s
(\Gamma \kappa) -q_{st} \right] \, ,
\label{e:linz}
\end{eqnarray}
where $n_x$ is the $x$ component of the unit vector
${\bf n}$ normal to the step. The normal step velocity is written as:
\begin{eqnarray}
V_n=\phi {\partial_t R
\over \left[1+(\partial_\theta R/R)^2 \right]^{1/2}} \, ,
\label{e:norm_v_pol}
\end{eqnarray}
and the step curvature
is defined with the sign convention:
\begin{eqnarray}
\kappa =\phi {R^{-1}+\partial_{\theta\theta}R^{-1}
\over \left[1+(\partial_\theta R/R)^2 \right]^{3/2}}\, .
\label{e:curv_pol}
\end{eqnarray}
We then linearize Eq.\ (\ref{e:linz}) for small
deformation of the island  $\rho(\theta)$.
With the Fourier transform of $\rho(\theta,t)$
as defined by Eq.\ (\ref{e:fourier}),
Eq.\ (\ref{e:linz}) takes the form:
\begin{eqnarray}
\lefteqn{\left[ i \omega \tau + n^2 (n^2-1)\right] \rho_{n \omega}}
\nonumber \\
&-& n \left[ (n+2) \rho_{n+1, \omega}+(n-2) \rho_{n-1, \omega} \right]
\phi\chi
={in\tau \over R_0}\phi q_{n\omega},
\label{e:pdlin}
\end{eqnarray}
where $\tau=a D_L \Gamma/R^4_0$. This equation implies
that $\rho_{0}$ is a constant; but this constant must vanish ($\rho_{0}=0$)
since $\rho$ is defined as the departure
from the mean radius $R_0$.
In Eq.\ (\ref{e:pdlin}) the eigenvalues
$i\omega$ do not depend on the sign of $\chi$ (as expected
physically). Thus, these eigenvalues also do not depend on $\phi$.
Hence, as stated above, atom and vacancy islands have the same
instability criterion $|\chi|>\chi_c$.

We first consider the case $\chi \ll 1$. We then expand step fluctuations
in the form $\rho_n=\rho_n^{(0)}+\chi\rho_n^{(1)}+\chi^2\rho_n^{(2)}+...$~.
To 0th order, we simply
get Eq.\ (\ref{e:pdlin}) without the term proportional to $\chi$.
The diffusion constant is calculated using the Fourier space relation:
\begin{eqnarray}
D_c=\langle |\rho_1|^2\rangle/t \; .
\end{eqnarray}
The equilibrium diffusion constant reads:
\begin{equation}
D_c^{eq}
={a^3 D_L \over \pi R_0^3} \; .
\label{e:dc_pd}
\end{equation}
To higher orders in $\chi$, Eq.~(\ref{e:pdlin}) shows the Langevin
force $q$ does not explicitly intervene, ultimately because $D_L$ is
supposed
to be independent of electromigration; hence,
\begin{eqnarray}
\lefteqn{\left[ i \omega \tau + n^2 (n^2-1)\right] \rho^{(m+1)}_{n \omega}}
\nonumber \\
&=& n \left[ (n+2) \rho^{(m)}_{n+1, \omega}
+(n-2) \rho^{(m)}_{n-1, \omega} \right] \phi\chi
\end{eqnarray}
Using this relation, we calculate the first correction
to the cluster diffusion constant:
\begin{equation}
D_c = D_c^{eq}\left( 1+\frac{\chi^2}{4} + O(\chi^4)\right) \; .
\end{equation}
The non-equilibrium cluster roughness can be calculated in a
similar way. We find:
\begin{equation}
W^2 = W^2_{eq}\left(1+b\chi^2 + O(\chi^4)\right),
\end{equation}
where $b \simeq 0.23$ is a numerical constant
resulting from infinite summations.
Since $\chi$ is of order unity, this correction can be non-negligible.
Note that the first correction to these quantities is proportional
to the electromigration force squared. Correspondingly,
$W$ and $D_c$ are invariant
under the inversion symmetry $F \rightarrow -F$.

Fluctuations increase with $F$.
In a linear theory, they
should diverge at the instability threshold.
If we define $\epsilon \equiv \chi_c-\chi$, then close to
the threshold, Eq.\ (\ref{e:pdlin}) can be written in terms of
the vector $\vec{\rho}=\{\rho_n\, : |n| \ge 2\}$
\begin{equation}
{\bf I}i\omega\vec{\rho}=
{\bf M}\vec{\rho}+\vec{b},
\end{equation}
where $\vec{b}$ is a noise term,
${\bf I}$ is the identity matrix and
the matrix ${\bf M}$ is expanded as ${\bf M}={\bf M_c}+\epsilon
{\bf M_1}$, where ${\bf M_c}$
has one null
eigenvalue (as can be seen, for example, in Ref.\ \onlinecite{wsh},
table II).
It is then easily shown that the squared roughness
\begin{eqnarray}
W^2 &=& \langle \vec{\rho}^T \vec{\rho} \rangle \nonumber \\
&=& \int {d\omega \over 2 \pi}\langle
\vec{b}^T(i\omega {\bf I}+{\bf M})^{-1T}
(-i\omega {\bf I}+{\bf M})^{-1}\vec{b} \rangle
\end{eqnarray}
diverges like $\epsilon^{-1}$ when $\epsilon \rightarrow 0$.
Hence, approaching the instability threshold,
\begin{equation}
W\sim\epsilon^{-1/2}.
\end{equation}
Note that $D_c$ does not
diverge: the instability concerns the morphology of the island,
not its motion.
The divergence of $W$ is analogous to
the case of an isolated step subject to a morphological
instability during growth,\cite{us} where the same exponent was found.
Close to threshold, fluctuations become large, and
nonlinear effects should be addressed, as shown in Ref.\onlinecite{km}
during growth.

\subsection{Monte Carlo simulations}
We here use kinetic Monte Carlo simulations to show that the
morphological instability of an island under periphery diffusion
leads to splitting.

We perform a 2D Monte Carlo simulation on
a square lattice. (Equivalently, the
simulation can be described as an SOS model in which
the height of the surface is either 0 or 1.)
The energy barrier for a move is taken to be proportional
to the number of {\it in-plane} nearest ($n$) and next-nearest ($n^{\prime}$)
neighbors of the atom before hopping.
Next-nearest neighbors are included to reduce the anisotropy of the steps.
Moves leading to an adatom (i.e. an atom with
no occupied nearest or next-nearest neighbor sites)
are forbidden. Atoms are allowed to hop only to nearest-neighbor sites.
Electromigration is taken into account as a direction-dependent bias in the
hopping barriers. The total barrier energy is then written:
\begin{eqnarray}
E_{ijk}=\varepsilon_{\rm d}(n_{ij}+n_{ij}^{\prime})+
\varepsilon_{\rm em}\cos([k-1]\pi/2-\theta_0),
\end{eqnarray}
where $i$ and $j$ are position coordinates along the $x$ and $y$ axis,
respectively;
$n_{ij}$ and $n_{ij}^{\prime}$ are the number of nearest and next-nearest
neighbors, respectively; $k=1,2,3,4$ is the direction of the move (to one
of the
nearest neighbors); and
$\varepsilon_{\rm d}$ and $\varepsilon_{\rm em}$ are the energies
associated with
diffusion and electromigration, respectively.
Explicitly, $\varepsilon_{\rm em}=F_{\rm st}a/2$;
$\theta_0$ is the angle between
of electromigration force and the [10] axis of the lattice.
Atoms and moves are picked randomly, and a move is performed
with probability proportional to $\exp(-E_{ijk}/k_BT)$.

A circular vacancy is chosen as the initial state.
The late stage of the
instability has been the subject of a recent
controversy.\cite{mb,Schim}
As for insulating voids in Ref.\onlinecite{Schim},
our atom and vacancy islands (equivalent to conducting voids)
split when they are unstable,
as illustrated in Fig.~3b.
For weaker electromigration (i.e. smaller $\chi$),
the cluster is stable, as shown in Fig.~3a.
In this figure, the steady state of the cluster at long time
is slightly elongated along the electromigration axis.
This elongation is probably a consequence of the anisotropy
of $D_L$ or $\tilde{\beta}$, which is not completely avoided
in the simulations.

Quantitatively the instability appears for radii larger than
$R_c \approx 13 a$ at $k_BT=0.6\varepsilon$, where the characteristic
energy $\varepsilon$ is the energy of a single bond. We note that an
elementary kink
costs energy $\varepsilon$ in our model. In the ``restricted"
approximation, in which
only elementary kinks (i.e. those repositioning the step by a single row)
are allowed,
the step stiffness becomes:
\begin{eqnarray}
\tilde{\beta}=(2+e^{\varepsilon/k_BT}){ k_BT \over 2a}\, .
\end{eqnarray}
Using this relation, we find $\tilde{\beta}=2.2\varepsilon/a$.
Now from the criterion $|\chi|>\chi_c$, we get
$R_0>R_c \approx 10.8a$, in good agreement with our simulations.
Note that the instability can also be used to determine
$\tilde{\beta}$ or $F$ when the other parameters are known.
{}From our simulation, with $R_c =13a$, we find
$\tilde{\beta}=3.3\varepsilon/a$.

As expected from the model, the same instability is found for
atom islands.

\begin{figure}
\centerline{
\epsfxsize=2.5in
\epsfbox{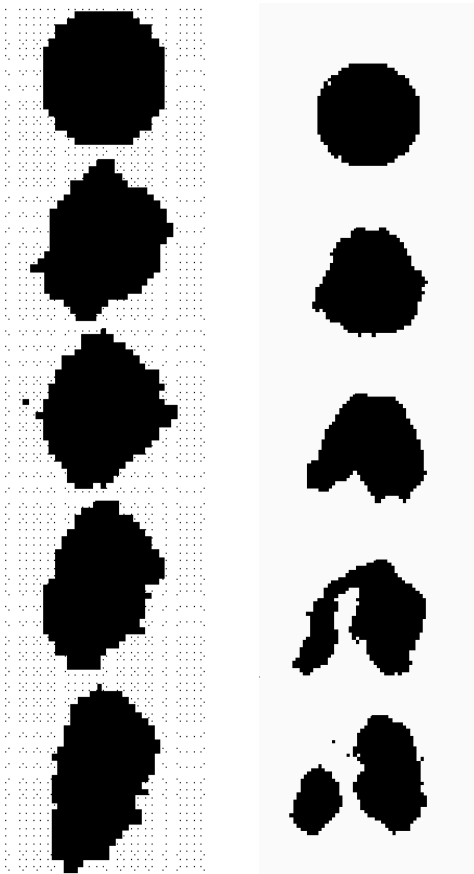}}
\caption{\it \protect \small
Monte Carlo simulations of a vacancy island in the PD
limit, with $F=0.1\varepsilon/a$ and $k_BT=0.6\varepsilon$.
a) Stable island, at $R_0=10a$.
b) Splitting, at $R_0=15a$.
}
\label{fig3}
\end{figure}

\section{2D transport regime}

\subsection{Steady states}
In the opposite limit of 2D transport,
the equilibrium behavior of these islands is well known.\cite{ke}
The cluster diffusion constant is
\begin{equation}
D_c^{eq} = \frac{ a^4Dc^0_{eq}}{\pi R_0} \left( \frac{1}{R_0+d_+}
+\frac{1}{R_0+d_-}\right) .
\label{e:dc_td}
\end{equation}
The corresponding scaling expectation of Eq.~(\ref{e:Vscale}) is
confirmed readily for weak electromigration, where we can work to first
order in
$\xi^{-1}$.  The islands remain circular and drift with velocity
\begin{equation}
\overline{V} = -\frac{a^2Dc^0_{eq}}{\xi} \left( \frac{R_0}{R_0+d_+}
-\frac{R_0}{R_0+d_-}\right) .
\label{e:Vterr}
\end{equation}

The first conclusion is that an Ehrlich-Schwoebel effect ($d_- > d_+$) is
needed for
non-zero drift.
We note in particular the remarkable prediction that for EC, the
drift velocity actually {\em increases} with increasing island size.
This behavior has been seen in Monte Carlo simulations by Wickham
and Sethna,\cite{ws} and leads to an exponential coarsening of
an assembly of clusters when coalescence is induced by drift-velocity
differences. Eq.\ (\ref{e:Vterr})
allows us to evaluate $d_+$ inside the voids of
Ref \onlinecite{ws}. Their findings lead to the expected result
that the length
$d_+$ increases as temperatures decreases. Moreover, their
phenomenological fitting expression for asymptotically large islands (cf.\
the caption of their Fig.\ 6),
$|\overline{V}|=v_0-v_1(a/R_0)$, can be retrieved from an expansion
of Eq.\ (\ref{e:Vterr}) for $R_0 \gg d_+$ (and $d_- \rightarrow \infty$).
Thus, we can identify their coefficients: $v_0 = a^2Dc^0_{eq}/\xi$ and $v_1 =
v_0d_+/a$.

Since Eq.\ (\ref{e:Vterr}) does not depend on $\phi$,
both atom and vacancy islands
drift in the {\it same} direction in the 2D transport regime.
Thus, we obtain a
very simple way to identify the dominant mass
transport mechanism, since atom and vacancy islands
drift in {\it opposite} directions in the PD limit. (Cf.\ Eq.\ (\ref{e:Ho}).)
Note also that when $d_->d_+$\ ---\ expected for the typical Ehrlich-Schwoebel
effect\ ---\ the direction of motion of the (adatom or vacancy) cluster in the 2D
transport regime is  opposite to the electromigration force.  (Cf.\ Fig.\ 2.)

For a vacancy or an atom cluster in the TD limit, the circular
steady state is an exact solution of the equations of motion
(even in the nonquasistatic limit).
As soon as the attachment is not instantaneous ($d_+ \ne 0$),
this is no longer true.
To second order in $\xi^{-1}$, a non-circular shape is found in steady
state for the ``interior" model (e.g. appropriate to a vacancy island with
an infinite Ehrlich-Schwoebel barrier) when $d_+ \ne 0$.
This shape is elongated perpendicular to the
electromigration force for vacancy islands (along the electromigration force
for atom islands), as can be understood from an intuitive argument:  For
small sticking probability, atoms detaching from the sides of the vacancy
island, will drift along the electromigration axis,
but will also have a residual drift toward the center of the vacancy if they
attach only after several attempts.

A macroscopic description, analogous to the
Bernoulli effect, provides an intuitive understanding
of this phenomenon. According to this effect, the pressure variation
perpendicularly to a flow is proportional to the kinetic energy
involved in this flow. Considering the flow of adatoms
inside the island, this relation takes the form:
\begin{equation}
J_{att} \times mV_{att} \propto {1\over 2}m c_{eq}
V_{adat}^2,
\label{e:JB}
\end{equation}
where the l.h.s. is the pressure variation and the r.h.s.
is the kinetic energy of the flow.
The adatom mass divides out,
it is irrelevant here because the dynamics is overdamped.
The first factor on the l.h.s.,
\begin{eqnarray}
J_{att}=\nu_+ (c-c_{eq}),
\label{b1}
\end{eqnarray}
is the flux of atoms attaching to the steps.
The second factor $mV_{att}$ is the momentum associated with
atoms attaching to the step, i.e., the product of the adatom mass $m$
and the macroscopic velocity
\begin{eqnarray}
V_{att}=D/(R_0+d_+)
\label{b2}
\end{eqnarray}
associated with
motion of an adatom across the island during its lifetime.
On the r.h.s. of Eq.\ (\ref{e:JB}), we use
$V_{adat}$, defined as the adatom drift
velocity needed to move the step at velocity $\overline{V}$. Hence,
\begin{eqnarray}
V_{adat}c_{eq} =\overline{V}/a^2 \;,
\label{b3}
\end{eqnarray}
where the cluster velocity $\overline{V}$ is given by Eq.\ (\ref{e:Vterr})
As a macroscopic property, $V_{adat}$ is different 
from the drift velocity $\langle v\rangle$ of
one adatom as defined in Eq.\ (\ref{e:einstein_1}).
Using Eqs.\ (\ref{b1}-\ref{b3})
in the Bernoulli relation Eq.\ (\ref{e:JB}),
we get the effective concentration change induced by the
electromigration force:
\begin{eqnarray}
c-c_{eq} \propto c_{eq}{R_0^2 \over \xi^2} {d_+ \over R_0+d_+} .
\label{e:bern}
\end{eqnarray}
Moreover, the concentration variation
associated with a change of local radius is known from
Gibbs-Thomson relation
\begin{eqnarray}
c-c_{eq} \propto c_{eq} \Gamma \delta R/R_0^2 \; .
\label{e:gt}
\end{eqnarray}
Balancing the destabilizing effect in Eq.\ (\ref{e:bern})
to the stabilizing one in Eq.\ (\ref{e:gt}), we get a
condition for a steady-state:
\begin{equation}
{\delta R \over R_0}
\propto {1 \over \Gamma}{R_0^2 \over \xi^2}
{d_+R_0 \over d_++R_0}.
\label{e:defber}
\end{equation}
Expanding our interior model to second order in $1/\xi$,
the island deformation involves only the
$n=2$ mode.  Defining
\begin{eqnarray}
\Delta \equiv {1 \over 2 R_0} [\rho(\pi/2) +\rho(-\pi/2) -\rho(0) -\rho(\pi)]
\; ,
\end{eqnarray}
we find,  quantitatively,
\begin{equation}
\Delta = \frac{1}{\Gamma}\frac{R_0^2}{6\xi^2}\frac{d_+R_0}{d_+ +R_0} \ ,
\label{delta_nd}
\end{equation}
in agreement with the evaluation in Eq.\ (\ref{e:defber}) based on the
Bernoulli
effect. In the EC limit, with $R_0 \sim 10^3a$ and $\xi \sim 10^5a$, one
finds that
$\Delta \sim 10^{-2}$. Thus, the deformation is small for an
electromigrating island
with interior dynamics.

When mass exchange with the exterior is allowed, there is no steady state.
To investigate further, we study the case of an (atom) island
exchanging matter only with the exterior, but with desorption allowed.
A term $F_{eq}-c/\tau_s$ is added to Eq.\ (\ref{e:mc_quas}).
Although the island is not stable in this case, we
further consider its behavior in order to analyze the tendency to
deviate from circularity as it shrinks.
To first order in $1/\xi$ we find a circular steady state with
velocity
\begin{eqnarray}
\overline{V}=-{1 \over \xi} \Omega D c_{eq}
{2x_sK_1 \over 2x_sK_1+d_+(K_2+K_0)}
\end{eqnarray}
where $K_n\equiv K_n(R_0/x_s)$ is the modified Bessel $K$ function and
$x_s=(D\tau_s)^{1/2}$. The $n=0$ mode is unstable, and the island has a
characteristic decay time
\begin{eqnarray}
\tau_{decay}={4 \pi R_0^2 \over \Omega D c_{eq}} \,\,
{x_sK_0+d_+K_1 \over K_1} \ .
\end{eqnarray}
To second order in $\xi^{-1}$ the island shape is not circular anymore, and
the resulting deformation reads:
\begin{eqnarray}
\Delta=-{d_+\over 3\Gamma}{R_0x_s \over 8 \xi^2}
{2x_s \over K_1+K_3}
{R_0 (K_2-K_1K_3)-x_sK_2K_1 \over x_sK_1+d_+(K_0+K_2)/2}
\end{eqnarray}
Since $\Delta$ is negative,
the elongation of an atom island is now along the electromigration axis.

This can also be qualitatively understood from the Bernoulli
effect, and the origin of elongation along the electromigration
axis can be traced back to a change of sign of the curvature
in the Gibbs-Thomson
relation (\ref{e:gt}).
To adapt the local equilibrium at the step to the lowering
of the pressure on the sides of the island, the curvature
has to be locally decreased, so that the island elongates
along the electromigration axis. In the limit of small desorption,
we get:
\begin{eqnarray}
\Delta=-{1  \over 2 \Gamma}{R_0^2 \over \xi^2}
{x_sd_+ \over R_0 +d_+}
\end{eqnarray}
Note that $\Delta$ diverges when $x_s \rightarrow \infty$,
in agreement with the previous claim that there is no
steady state in this limit. We notice that this formula is
similar to Eq.\ (\ref{delta_nd}), with a factor of $R_0$ replaced
by the new cut-off length $x_s$.
Taking $\Gamma \sim a$, $d_+ \sim 10^4a$, $R_0 \sim 10^3a$, $x_s \sim10^6a$,
and $\xi \sim 10^8a$, we get $\Delta \sim 0.5$.
Hence, a large deformation
of the island can be found
in the exterior model.

\subsection{Validity of the quasistatic limit}

In this section,
we consider a vacancy island in the absence of electromigration
and in the non-quasistatic regime.
We use polar coordinates $r$ and $\theta$.
The $n$th component of the Fourier transform of the concentration
with respect to $\theta$ is expanded as
\begin{eqnarray}
c_n(r,t)=\sum_{m=0}^{\infty}a_{nm}(t)r^m \ .
\label{e:exp_conc}
\end{eqnarray}
Using this expression in the interior model,
we look for solutions of the
form $a_{nm}(t)={\rm exp}(i\omega t)a_{nm}$.
By symmetry $a_{nm}=a_{-nm}$, and from analyticity
of $c$ at $r=0$, $a_{nm}=0$ when $n-m$ is odd.
Using the recursion equations resulting from
Eqs.(\ref{e:mc_ter}-\ref{e:dtc}), it is the possible to show
that $a_{nm}=0$ when $m \le |n|$, and that all $a_{nm}$ can
be calculated from the free parameters $a_{n|n|}$.
We thus calculate the $a_{nm}$ as a function of the $a_{n|n|}$,
and plug this result into the boundary conditions
Eqs.\ (\ref{e:lin_kin}-\ref{e:conslaw})
--with $J_{st}=0$.
We find
the non-quasistatic dispersion relation
from the requirement that the
prefactors of the $a_{n|n|}$'s should cancel:
\begin{eqnarray}
\lambda&=& -a^2\tilde{c}_{eq}{\Gamma \over R_0}(n^2-1)
\times
\nonumber\\
& \times &
\left[{d_+ \over R_0} +\left(|n|
+\lambda^{1/2}{I_{|n|+1}(\lambda^{1/2}) \over
I_{|n|}(\lambda^{1/2})}\right)^{-1}\right]^{-1} \; ,
\label{e:nonquas}
\end{eqnarray}
where $I_n$ is the modified
Bessel $I$ function, and $\tilde{c}_{eq}=c_{eq}^0{\rm exp}(-\Gamma/R_0)$.
Since $\lambda\equiv i\omega R_0^2/D$,
$\Re{\em e}[ \lambda]$ is proportional to the growth rate
of the perturbations.
In the quasistatic limit, Eq.\ (\ref{e:mc_quas})
is solved on the interior terrace instead of Eq.\ (\ref{e:mc_ter}).
We then find the following dipersion relation:
\begin{eqnarray}
\lambda=-a^2\tilde{c}_{eq}{\Gamma \over R_0}(n^2-1)
\left[{d_+ \over R_0} +|n|^{-1}
\right]^{-1}
\label{e:quas}
\end{eqnarray}

\noindent
Note that in the quasistatic limit, the $n=0$ mode is
frozen (i.e. $\lambda=0$ when $n=0$).
Moreover, we obtain here the standard TD (or EC) limit
by letting $d_+\rightarrow 0$ (or $d_+\rightarrow \infty$),
leading to $i\omega \sim R_0^{-2}$ (or $\sim R_0^{-3}$).
{}From Eq.\ (\ref{e:nonquas}), the quasistatic limit
Eq.\ (\ref{e:quas}) is recovered
when $|\lambda| \ll n^2$.
Using the quasistatic expression for
$\lambda$, this condition can be rewritten:
\begin{equation}
\Gamma a^2 \tilde{c}_{eq}\nu_+ \ll {R_0 \nu_+ \over |n|}+D \; ,
\label{e:quascond}
\end{equation}
where $|n|>1$.
For the approximation to be valid for all modes,
it is sufficient to satisfy the stronger condition
\begin{eqnarray}
\Gamma a^2 \tilde{c}_{eq}\nu_+ \ll D \, .
\label{e:quascrit}
\end{eqnarray}
This inequality should be interpreted as follows:
the effective diffusion constant of the step (l.h.s.) must be
much smaller than that of the adatoms on the terrace (r.h.s.).

The mode-dependent term on the  r.h.s. of Eq.\ (\ref{e:quascond}), related to
attachment-detachment, indicates that the
quasistatic limit is more difficult to achieve at
large $|n|$, i.e. small wavelengths, 
because these modes have shorter relaxation times.
Using a small-wavelength cut-off 
$|n|^*\sim R_0/a$ on the r.h.s. of
Eq.\ (\ref{e:quascond}), we see that
Eq.\ (\ref{e:quascrit}) is valid in the limit of slow kinetics,
$\nu_+ \ll D/a$.
Conversely, when $\nu_+\gg D/a$, we find
\begin{eqnarray}
a\Gamma\tilde{c}_{eq} \ll 1 \, ,
\label{e:quas_td}
\end{eqnarray}
which is the criterion for the quasistatic limit to be valid
in the fast-kinetics (i.e. TD) regime. 

As an example, let us consider the case of Si(111) at $T \sim 1000$K.
{}From Ref. \onlinecite{will}, we have
$a^2 c_{eq}< 0.1$ and $d_+=D/\nu_+\sim 10^4a$, and from
Ref. \onlinecite{metois},
$\Gamma\sim 10 {\rm \AA}$. Eq.\ (\ref{e:quascrit})
is then easily checked to be valid: the quasistatic 
approximation can be used.

\subsection{Stability analysis}

We now go back to the electromigration problem,
and perform a linear stability analysis of the
non-quasistatic model equations.
We consider the TD limit ($d_+=0$) of the interior model.
We perform a stability analysis of perturbations with
respect to the steady state with constant concentration $c=\tilde{c}_{eq}$
and circular shape $R=R_0$. To do so, we again expand
the concentration using Eq.\ (\ref{e:exp_conc}), now in Eqs.\
(\ref{e:mc_ter}-\ref{e:dtc}), and seek solutions of the  form $a_{nm}(t)={\rm
exp}(i\omega t)a_{nm}$.  Matching coefficients for
for each $r^m$ leads to a set of recursion equations for the $a_{nm}$'s
for $m \ge 2$:
\begin{eqnarray}
{i\omega \over D} a_{n,m-2}= -a_{nm}(n^2-m^2)
&-&{1 \over 2 \xi}\Big[(n+m)a_{n+1,m+1}
\nonumber \\
&-&(n-m)a_{n-1,m-1}\Big],
\end{eqnarray}
supplemented by the relations
\begin{eqnarray}
a_{n0}&=& 0 \;{\rm for}\; n \neq 0 \; ,
\nonumber \\
a_{n1}&=& 0\;{\rm for}\; |n| \neq 1 \; .
\end{eqnarray}
Boundary conditions Eqs.\ (\ref{e:lin_kin}-\ref{e:conslaw})
then provide two set of relations for the $a_{nm}$'s and the $\rho_n$'s.
When the $\rho_n$'s are eliminated,
a set of relations among the $a_{nm}$'s is found:
\begin{eqnarray}
0=\sum_{m=0}^{\infty} \Big[
-{i\omega R_0^2 \over D} a_{nm}
-{a^2\tilde{c}_{eq} \over R_0}(n^2-1)\Gamma\Big\{ ma_{nm}
\nonumber \\
 -{R_0 \over 2 \xi}(a_{n+1,m}+a_{n-1,m}) \Big\} \Big] .
\label{e:disp}
\end{eqnarray}
Numerical solution of
this set of equations reveals an instability for $R_0/\xi > 0.1$ related to
the existence of a positive eigenvalue $i\omega$ in this range
of parameters.  It is interesting to note
that the instability threshold is independent of $\Gamma$.
To investigate the nature of the instabilities of these islands, we again
perform Monte Carlo simulations. Rather than the splitting seen in the PD
limit, we find a new type of instability.

\subsection{Monte Carlo Simulations}
The code used here is similar to that used for a vacancy island
in the PD limit,
but with adatoms now allowed inside. Moreover, in order to tune
attachment-detachment kinetics, the hopping
probability is decreased by a factor $p$, $0<p\le 1$, when the
number of neighbors changes during the hop.\cite{liu}

\begin{figure}
\centerline{
\epsfxsize=2.5in
\epsfbox{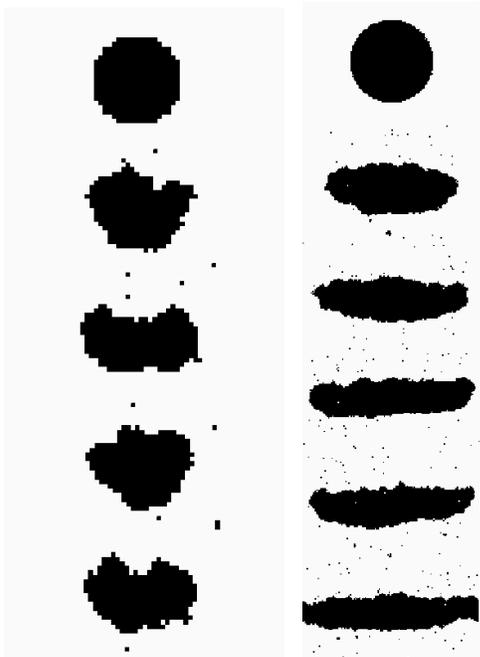}}
\caption{\it \protect \small
Monte Carlo simulations of a vacancy island in the TD
limit, with $F=0.1\varepsilon/a$, $k_BT=0.6\varepsilon$.
a) ``Stable" island, at $R_0=10a$.
b) Slit formation, for $R_0=30a$.
}
\label{fig4}
\end{figure}

At high temperature $k_BT \geq 0.7\varepsilon$,
micro-vacancies escape noticeably from the steps. This
might heal possible instabilities by producing
a micro-vacancy where a cusp is present.
We therefore chose to perform our simulations at $k_BT=0.6\varepsilon$.
Even though a few micro-vacancies still escape from the clusters at
this temperature, they are rare enough so that their effect on the dynamics
can be considered negligible.

No drastic change in the island shape is seen at the threshold given by
the linear stability analysis. Let us define the front and the back
side of the island with respect to cluster motion as the first
and the last part passing at a given value of $x$.
When electromigration is increased, the front side
is stabilized. The back side is
destabilized and exhibits chaotic behavior,
but its roughness remains finite,
as shown in Fig.~4a.
We conjecture that this is the instability that
we found from linear analysis.
Although some instability
is seen in this regime, we classify the island as stable because
its shape is not affected drastically.

Increasing the electromigration force, we see that
unstable slit-like shapes appear, as depicted in Fig.~4b.
Since these shapes are seen for all orientations of the electromigration
force, they are not due to lattice anisotropy.
This instability contrasts qualitatively with the splitting
found in the PD regime (see Fig.~3b).

Changing the value of $p$ does not seem to change this scenario,
but we have not performed simulations in the limit where $p$ is
small ($p<0.1$) due to the large computation time needed there.
Numerical solution of the model equations would be needed
to understand in greater detail these different regimes.

\section{Electromigration bias in attachment-detachment}

Finally, we introduce a possible electro-bias between attachment and
detachment.  Extracting one atom from the step, not counting any bond
breaking, requires work $a${\bf F}$_{\rm em}\cdot {\bf \hat{n}}$ along the
normal to the step.\cite{note1}
{}From a linearized Gibbs-Thomson relation, we
find that the effective equilibrium concentration becomes
\begin{eqnarray}
c_{eq}= c^0_{eq}(1 + a{\bf F}_{\rm em}{\bf \cdot \hat{n}}/k_BT) \; .
\end{eqnarray}
Defining
$\xi_{\pm}^{-1} = {\bf F}_{\rm em}{\bf \cdot \hat{n}}/k_BT$ on the lower or
upper side of the boundary step edge, respectively, we find that electro-bias
contributes to the velocity of Eq.~(\ref{e:Vterr}) an extra term

\begin{equation}
\delta\overline{V} = a^3Dc^0_{eq} \left(
\frac{\xi^{-1}_{1,+}}{R_0+d_+}
-\frac{\xi^{-1}_{1,-}}{R_0+d_-}\right) \ ,
\label{e:Veb}
\end{equation}

\noindent where the subscript indicates the $n=1$ component of the Fourier
transform of $1/\xi_\pm(\theta)$.
When there is no Ehrlich-Schwoebel effect ($d_+ =
d_-$), $\overline{V}$ from Eq.~(\ref{e:Vterr})
vanishes, and the drift velocity
is only due to this new contribution combined with PD effects. In the EC limit,
$\delta\overline{V}$ does not depend on the size of the cluster, while in the
TD limit $\delta\overline{V} \sim 1/R_0$. This is the same size dependence as
that associated with the PD limit.
Since $\delta\overline{V}$ decreases with increasing island size,
we expect this contribution to be small for large islands.

The deformation of the cluster is proportional to $1/\xi$,
rather than $1/\xi^2$ as seen in the other regimes. Thus electrobias might
induce larger deformations of the clusters. However, since we
do not know the angular dependence of these forces, we cannot
go further in the analysis.

\section{Summary and perspectives}

We have studied the response of single-layer clusters to electromigration
in different regimes associated with different mass-transport mechanisms.
Some of the main results are collected in table 1.
The drift velocity and shape changes agree
with those found in earlier published
calculations,\cite{Ho} and give a theoretical explanation
and explicit expressions for those found with Monte Carlo simulations.

The instability is found to
depend qualititatively on the mass transport mechanism. PD is
associated with splitting, and 2D transport with slit formation.
Although the instability threshold of the latter
may be beyond the experimental range for 2D islands, it
could well be important in the case of disordered oxygen domains
in YBaCuO,\cite{mlb,ws}
and for void dynamics,\cite{Ho,mb,wsh,Schim} where higher
current densities are used. In the latter case, 2D transport might occur
on the substrate across the void.

Fluctuations are found to be affected by electromigration.
The non-equilibrium cluster diffusion constant and roughness
were studied. The island roughness is found to diverge
as the instability threshold is approached.

Non-linear analysis is needed for a better understanding
of these instabilities.
Moreover, cross-over regimes should be studied in order to
catalogue  completely the behavior of these islands.
For a more realistic description of the islands,
especially at low temperature, anisotropy should also be
included.\cite{ws,Schim,Gung} Indeed, it was already shown experimentally that
extremely anisotropic properties (such as diffusion) could lead to
very different behavior.\cite{metpimp}

\begin{table}
\begin{tabular} {l||c|c|c}
 & {\it PD} & {\it TD} & {\it EC (AD)}  \\  \hline \hline
Length & & & \\
\hspace{2mm} criterion & \raisebox{1.5ex}[0pt]{$R_0^2 \ll \ell_c(R_0+d)$} &
\raisebox{1.5ex}[0pt]{$R_0 \gg d_+,d_-$} & \raisebox{1.5ex}[0pt]{$R_0 \ll
d_+,d_-$} \\ \hline
& & & \\
\raisebox{1.5ex}[0pt]{$\langle V(R_0) \rangle$} &
\raisebox{1.5ex}[0pt]{$1/R_0$} & \raisebox{1.5ex}[0pt]{$1$} &
\raisebox{1.5ex}[0pt]{$R_0$} \\
\hline Atom/vacancy & & \multicolumn{2}{c}{} \\
\hspace{2mm}direction & \raisebox{1.5ex}[0pt]{Opposite} &
\multicolumn{2}{c}{\raisebox{1.5ex}[0pt]{Same}} \\
\hline Steady state  & \multicolumn{2}{c|}{}  & None \\ \cline{4-4}
\hspace{2mm} shape & \multicolumn{2}{c|}{\raisebox{1.5ex}[0pt]{Circular}} &
Non-circ. \\
\hline
Instability &  & \multicolumn{2}{c} {``Slit" $\parallel$ {\bf F}} \\
\cline{3-4}
\hspace{2mm} morphology & \raisebox{1.5ex}[0pt]{Splitting}
&  \multicolumn{2}{c} {Slit $\perp$ {\bf F}} \\ \hline
& & \multicolumn{2}{c}{} \\
\raisebox{1.5ex}[0pt]{$D_c$} & \raisebox{1.5ex}[0pt]{$D_c^{eq} (1 +
\chi^2/4)$} &
\multicolumn{2}{c}{\raisebox{1.5ex}[0pt]{Anisotropic}} \\
\end{tabular}
\caption[tabl1]{Summary of results. In general the entries apply to both
atom and
vacancy islands.  The criteria for the three regimes in terms of the
characteristic lengths are given in the first row.  The inequality for PD is
dominant; when it holds, one has PD even if the inequality for TD or EC is
satisfied.  The characteristic length $\ell_c$ is
$D_L/aDc_{eq}^0$, with equilibrium values for the parameters.
For the
third row, the relative direction of the drift velocity of atom and vacancy
islands is tabulated.  In the fourth and fifth rows, the upper (lower) entry
is for atom/exterior (vacancy/interior) islands.
PD, TD, and EC indicate the mode of mass
transport: periphery
diffusion, terrace diffusion, and evaporation-condensation, respectively.
Reidentifying the letter D, the last is often called AD (attachment-detachment)
to indicate that the process involves just a 2D ``gas" on the terraces.}
\end{table}

This study is a step toward the understanding of the evolution
of more complex surface geometries in the presence of a diffusion drift,
such as 3D islands or groves or assemblies of clusters.
To complete this exploration, the fluctuations
in the 2D transport regime should be studied. It would also be useful to check
with Monte Carlo simulations the predictions about the diffusion constant
and the
divergence of morphological fluctuations.

We note in closing that, as the surface is affected
by electromigration, so electromigration itself is affected
by the morphology of the surface. Step shadowing,
adatom crowding, or surface roughness induce changes
in the local electromigration force.\cite{rous_step}
A quantitative description of experiments should take these feedback
mechanisms into account.

Aknowledgements:
This work was supported by NSF MRSEC grant DMR-96-32521.  We acknowledge
helpful conversations with  E.D. Williams and P.J. Rous,
and thank J. Krug for comments on the manuscript.


\begin{thebibliography}{99}
\bibitem{lask} A.V. Latyshev, A.L. Aseev, A.B. Krasilnikov,
and S.I. Stenin, Surf. Sci. {\bf 213}, 157 (1989).

\bibitem{Ho} P.S. Ho, J. Appl. Phys. {\bf 41}, 64 (1970).

\bibitem{mlb} B.H. Moeckly, D.K. Lathrop, and R.A. Buhrman,
Phys. Rev. B {\bf 47}, 400 (1993).

\bibitem{ws} 
L.K. Wickham and J.P. Sethna, Phys. Rev. B {\bf 51}, 15017 (1995).

\bibitem{sz} B. Schmittmann and R.K.P. Zia, {\it Statistical Mechanics of
Driven Diffusive Systems} [Phase Transitions and Critical Phenomena, vol.
17, C. Domb
and J.L. Lebowitz, Eds.] (Academic, London, 1995).

\bibitem{Rosen} G. Rosenfeld, K. Morgenstern, and G. Comsa, in {\it Surface
Diffusion: Atomistic and Collective Processes} [NATO ASI series B 360], M.C.
Tringides, Ed. (Plenum, New York, 1997), 361.


\bibitem{controversy3D} L. Turban, P. Nozi\`eres, and M. Gerl,
J. Phys. (Paris) {\bf 37}, 159 (1976);
A. Lodder, Sol. St. Comm. {\bf 79}, 143 (1991).

\bibitem{kk} D. Kandel and E. Kaxiras, Phys.\ Rev.\ Lett.\ {\bf 76}, 1114
(1996).

\bibitem{Ish} H. Ishida, Phys. Rev. B {\bf 49}, 14610 (1994).

\bibitem{rew} P.J. Rous, T.L. Einstein, and E.D. Williams,
Surf. Sci. {\bf 315}, L995 (1994).

\bibitem{Rous} P.J. Rous, Phys. Rev. B {\bf 59}, 7719 (1999).

\bibitem{sc/} E.S. Fu, D.-J. Liu, M.D. Johnson, J.D. Weeks, and E.D. Williams,
Surf.\ Sci.\ {\bf 385}, 259 (1997);
J.-J. M\'etois and M. Audiffren,
Int. J. Mod. Phys. B {\bf 11}, 3691 (1997).

\bibitem{ke} a) S.V. Khare, N.C. Bartelt, and T.L. Einstein,
Phys. Rev. Lett. {\bf 75}, 2148 (1995); b) S.V. Khare and T.L.
Einstein, Phys. Rev. B {\bf 54}, 11752 (1996);
c) {\bf 57}, 4782 (1998) d) in Dynamics of Crystal Surfaces and Interfaces,
P.M. Duxbury and T. Pence, eds. (Plenum, New York, 1997), 83.

\bibitem{stoy} S. Stoyanov, Jpn. J. Appl. Phys. {\bf 29},
L659 (1990).

\bibitem{hkgibh} J.B. Hannon, C. Kl\"unker, M. Giesen, H. Ibach, N.C. Bartelt,
and J.C. Hamilton, Phys. Rev. Lett. {\bf 79}, 2506 (1997).

\bibitem{hkma} J. Heinonen, I. Koponen, J. Merikoski, and T. Ala-Nissila, Phys.
Rev. Lett. {\bf 82}, 2733 (1999).

\bibitem{oplcm} O. Pierre-Louis and C. Misbah,
Phys. Rev. Lett. {\bf 76}, 4761 (1996).

\bibitem{vil_pimp} J. Villain and A. Pimpinelli, {\it
Physique de la Croissance Cristalline}, french edition only,
(Eyrolles Alea Saclay, Paris, 1995).

\bibitem{kex} In Eq.\ (6) of ref.\ \onlinecite{ke}a and Eq.\ (17) of ref.\
\onlinecite{ke}b, the numerical factor in the denominator was written
as $n^2$ rather than $n^2 -1$. Their results are not affected
by this change.

\bibitem{ost} P.W. Voorhees, J. Stat. Phys. {\bf 38}, 231 (1985); Annu.
Rev. Mater. Sci. {\bf 22}, 197 (1992).

\bibitem{smol} D.S. Sholl and R.T. Skodje, Phys. Rev. Lett. {\bf 75}, 3158
(1995); C.R. Stoldt, C.J. Jenks, P.A. Thiel, A.M. Cadilhe, and J.W. Evans, J.
Chem. Phys. {\bf 111}, 5157 (1999).

\bibitem{mrpc} K. Morgenstern, G. Rosenfeld, B. Poelsema, and G. Comsa,
Phys. Rev. Lett. {\bf 74}, 2058 (1995).

\bibitem{note_lc} In the notation of Ref. 12 b,d,
$\ell_c=R_{st}^2/R_{su}$.

\bibitem{grub} E. E. Gruber,
J. Appl. Phys. {\bf 38}, 243 (1967).

\bibitem{mbmk} H. Mehl, O. Biham, O. Millo, and M. Karimi, Phys. Rev. B
{\bf 61}, 4975 (2000).

\bibitem{mb}M. Mahadevan and R.M. Bradley,
J. Appl. Phys. {\bf 79}, 6840 (1996).
\bibitem{wsh}
W. Wang, Z. Suo, and T.-H. Hao,
J. Appl. Phys. {\bf 79}, 2394 (1996).

\bibitem{Schim}  M. Schimschak and J. Krug,
Phys. Rev. Lett. {\bf 80}, 1674 (1998);
J. Appl. Phys. {\bf 87}, 695 (2000).

\bibitem{us} M. Uwaha and Y. Saito,
Phys. Rev. Lett. {\bf 68}, 224 (1992).

\bibitem{km} A. Karma and C. Misbah, Phys. Rev. Lett.
{\bf 71}, 3810 (1993).


\bibitem{will} H.-C. Jeong, E.D. Williams,
Surf. Sci. Rep. {\bf 34}, 171 (1999).

\bibitem{metois} C. Alfonso, J.M. Bermond, J.C. Heyraud,
and J.J. M\'etois, J. Phys. I (France), {\bf 5}, 443 (1995).

\bibitem{liu} D.-J. Liu and J.D. Weeks, Phys. Rev. B {\bf 57}, 14891 (1998).

\bibitem{note1} For greater accuracy, we should integrate along
the detachment path.  Note that the force depends on the orientation of the
step with respect to {\bf F}.

\bibitem{Gung} M.R. Gungor and D. Maroudas,
Appl. Phys. Lett. {\bf 72}, 3452 (1998).

\bibitem{metpimp} J.-J. M\'etois, J.-C. Heyraud, and A. Pimpinelli,
Surf. Sci. {\bf 420}, 250 (1999).

\bibitem{rous_step} P.J. Rous, unpublished.

\end{thebibliography}
\end{document}